\documentclass[aps,prl,showpacs,twocolumn]{revtex4}
\usepackage{bm,color,amsmath,amssymb,mathrsfs,latexsym,graphicx,psfrag}

\newcommand{\ket}[1]{\left|#1\right\rangle}
\newcommand{\bigket}[1]{\bigl|#1\bigr\rangle}
\newcommand{\textket}[1]{|#1\rangle}

\newcommand{\up}{\uparrow}
\newcommand{\dw}{\downarrow}

\newcommand{\vac}{\left|\,0\,\right\rangle}

\def\ie{{\it i.e.},\ }

\begin{document}
\title{Non-Abelian Statistics in a Quantum Antiferromagnet}
\author{Martin Greiter and Ronny Thomale} \affiliation{Institut f\"ur
  Theorie der Kondensierten Materie, Universit\"at Karlsruhe, D 76128
  Karlsruhe} \pagestyle{plain}

\date{\today}

\begin{abstract}
  We propose a novel spin liquid state for a spin $S=1$
  antiferromagnet in two dimensions.  The ground state violates P and
  T, is a spin-singlet, and is fully invariant under the lattice
  symmetries.  The spinon and holon excitations are deconfined and
  obey non-abelian statistics.  We present preliminary numerical
  evidence that the universality class of this topological liquid can
  be stabilized by a local Hamiltonian involving three-spin
  interactions.  We conjecture that spinons in spin liquids with spin
  larger than $1/2$ obey non-abelian statistics in general.
\end{abstract}


\maketitle

\emph{Introduction}---Fractional quantization in two-di\-men\-sio\-nal
quantum liquids is witnessing a renaissance of interest in present
times.  The field started about a quarter of a century ago with the
discovery of the fractional quantum Hall effect, which was explained
by Laughlin~\cite{laughlin83prl1395} in terms of an incompressible
quantum liquid supporting fractionally charged (vortex or)
quasiparticle excitations.  When formulating a hierarchy of quantized
Hall states~\cite{haldane83prl605,halperin84prl1583
} to
explain the observation of quantized Hall states at other filling
fractions fractions, Halperin~\cite{halperin84prl1583} noted that
these excitations obey fractional statistics~\cite{Wilczek90}, and are
hence conceptually similar to the charge-flux tube composites
introduced by Wilczek two years earlier~\cite{wilczek82prl957}.

The interest was renewed a few years later, when
Anderson~\cite{Anderson87s1196} proposed that hole-doped Mott
insulators, and in particular those described by the $t$--$J$ model
universally believed to describe the CuO planes in high $T_\text{c}$
superconductors, can be described in terms of a spin liquid (\ie a
state with strong, local antiferromagnetic correlations but without
long range order), which would likewise support fractionally quantized
excitations.  In this proposal, the excitations are spinons and
holons, which carry spin 1/2 and no charge or no spin and charge $+e$,
respectively.  The fractional quantum number of the spinon is the
spin, which is half integer while the Hilbert space (for the undoped
system) is built up of spin flips, which carry spin one.  One of the
earliest proposals for a spin liquid supporting deconfined spinon and
holon excitations is the (abelian) chiral spin liquid (CSL).
Following up on an idea by D.H.~Lee, Kalmeyer and
Laughlin~\cite{Kalmeyer-87prl2095} proposed that a quantized Hall wave
function for bosons could be used to describe the amplitudes for
spin-flips on a lattice.  The CSL state did not turn out to be
relevant to CuO superconductivity, but remains one of very few
examples of two-dimensional spin liquids with fractional quantization.
The other established examples are the resonating valence bond (RVB)
phases of the Rokhsar-Kivelson model~\cite{kivelson-87prb8865} on the
triangular lattice identified by Moessner and 
Sondhi~\cite{moessner-01prl1881} and of the Kitaev
model~\cite{Kitaev06ap2}.
 
The present renaissance of interest in fractional quantization is due
to possible applications of states supporting excitations with
\emph{non-abelian statistics} to the rapidly evolving field of quantum
computation and cryptography. 
The paradigm for this universality class is the Pfaffian state
introduced by Moore and Read~\cite{Moore-91npb362} in 1991.  The state
was proposed to be realized at the experimentally observed fraction
$\nu =5/2$~\cite{ Willet-87prl1776} (\ie at $\nu =1/2$ in the second
Landau level) by Wen, Wilczek, and one of us~\cite{greiter-92npb567},
a proposal which recently received experimental support through the
direct measurement of the quasiparticle charge~\cite{dolev-08n829,radu-08s899}.
Pfaffian type states are further conjectured to be realized for
one-dimensional bosons with three-body hard core interactions in
general~\cite{paredes-07pra053611}.  The Moore--Read state possesses
$p+ip$ wave pairing correlations.  The flux quantum of the vortices is
one half of the Dirac quantum, which implies a quasiparticle charge of
$e/4$.  Like the vortices in a $p$ wave superfluid, these
quasiparticles possess Majorana-fermion states~\cite{Read-00prb10267}
at zero energy (\ie one fermion state per pair of vortices, which can
be occupied or unoccupied).  A Pfaffian state with $2L$ spatially
separated quasiparticle excitations is hence $2^L$ fold degenerate, in
accordance with the dimension of the internal space spanned by the
zero energy states.  While adiabatic interchanges of quasiparticles
yield only overall phases in abelian quantized Hall states, braiding
of half vortices of the Pfaffian state will in general yield
non-trivial changes in the occupations of the zero energy
states~\cite{Ivanov01prl268,Stern-04prb205338}, which render the
interchanges non-commutative or non-abelian.  In particular, the
internal state vector is insensitive to local perturbations---it can
\emph{only} be manipulated through braiding of the vortices.
These properties together render
non-abelions preeminently suited for applications as protected qubits
in quantum computation
~\cite{nayak-08rmp1083}.  Non-abelian
anyons further appear in certain other quantum Hall states 
including the Read-Rezayi states~\cite{Read-99prb8084}, in the
Kitaev model~\cite{Kitaev06ap2}, and in the
Yao-Kivelson model~\cite{yao-07prl247203}.

In this Letter, we propose a novel chiral spin liquid state for an
$S=1$ antiferromagnet.  The spinon and holon excitations of this state
are deconfined and obey non-abelian statistics, with the braiding
governed by Majorana fermion states.  The state violates time reversal
(T) and parity (P), is a spin singlet, can be formulated on any
lattice type, and fully respects all the lattice symmetries.  The
state possesses a 3-fold topological degeneracy on the torus geometry.
We provide preliminary numerical evidence that the state can be
stabilized on the triangular lattice by a local Hamiltonian involving
three-spin interactions.  Finally, we conjecture that spinons in spin
liquids with spin larger than 1/2 might obey non-abelian statistics in
general.
%

\emph{Non-abelian chiral spin liquid state}---The state we propose is
most easily written down for a circular droplet with open boundary
conditions occupying $N$ sites of a triangular or square lattice $S=1$
antiferromagnet.  The wave function for re-normalized spin flips,
\begin{equation}
  \label{eq:nacsl}
  \psi_0[z_i]=\text{Pf}\left(\frac{1}{z_{j}-z_{k}}\right)
  \prod_{i<j}^{N}(z_i-z_j)\prod_{i=1}^{N}\,G(z_i)\,e^{-\frac{\pi}{2}|z_i|^2}
\end{equation}
is given by a bosonic Pfaffian state in the complex
coordinates $z\equiv x+iy$ supplemented by a gauge phase
$G(\eta_\alpha)$.  The Pfaffian is given by the fully
antisymmetrized sum over all possible pairings of the $N$ coordinates,
\begin{equation}
  \label{eq:pfaff}
  \text{Pf}\left(\frac{1}{z_i -z_j}\right)\equiv
  \mathcal{A}
  \left\{
    \frac{1}{z_1-z_2}\cdot\,\ldots\,\cdot\frac{1}{z_{N-1}-z_{N}}
  \right\}.
\end{equation}
The ``particles'' $z_i$ represent re-normalized spin flips acting on a
vacuum with all spins in the $S^{\rm{z}}=-1$ state,
\begin{equation}
  \label{eq:nacslket}
  \ket{\psi_0}=\sum_{\{z_1,\dots,z_{N}\}} 
  \psi_0(z_1,\dots,z_N)\
  \tilde{S}_{z_1}^{+}\dots\tilde{S}_{z_{N}}^{+}\, 
  \ket{-1}_N,
\end{equation}
where the sum extends over all possibilities of distributing the 
$N$ ``particles'' over the $N$ lattice sites allowing for double 
occupation, and
\begin{equation}
  \label{eq:spinflip}
  \tilde{S}_{\alpha}^{+} \equiv\frac{S^{\rm{z}}_{\alpha}+1}{2} S_\alpha^{+},\quad
  \ket{-1}_N\equiv\otimes_{\alpha=1}^N \ket{1,-1}_{\alpha}.
\end{equation}
The lattice may be anisotropic; we have chosen the lattice constants
such that the area of the unit cell spanned by the primitive lattice
vectors is set to unity.  For a triangular or square lattice with
lattice positions given by $\eta_{n,m}=na+mb$,
where $a$ and $b$ are the primitive lattice vectors in the complex
plane, 
$G(\eta_{n,m})= (-1)^{(n+1)(m+1)}$~\cite{zou-88prb11424,
Kalmeyer-87prl2095}.

\vspace{3pt} \emph{Singlet property}---While the topological
properties, and in particular the non-abelian statistics of the
fractionalized excitations of \eqref{eq:nacsl}, are suggestive to
those familiar with Pfaffian states, the invariance under spin
rotation and lattice symmetries is less so.  We content ourselves here
with a direct proof of the singlet property, which at the same time
serves to motivate the necessity for the re-normalization of the
spin-flip operators \eqref{eq:spinflip}.

Since $S^z_\text{tot}\ket{\psi_0}=0$ by construction, it is sufficient
to show $S^-_\text{tot}\ket{\psi_0}=0$.  Note first that when we
substitute \eqref{eq:nacsl} with \eqref{eq:pfaff} into
\eqref{eq:nacslket}, we may omit the antisymmetrization $\mathcal{A}$
in \eqref{eq:pfaff}, as it is taken care by the commutativity of the
bosonic operators $\tilde{S}_{\alpha}$.  (Throughout this Letter, we
do not keep track of overall normalization factors.)  Let
$\tilde\psi_0$ be $\psi_0$ without the operator $\mathcal{A}$ in
\eqref{eq:pfaff}.
%
Since $\tilde\psi_0[z_i]$ is still symmetric under interchange of pairs, we
may assume that a spin flip operator $S^-_\alpha$ acting on
$\textket{\tilde\psi_0}$ will act on the pair $(z_1,z_2)$:
\begin{equation}
  \label{eq:salphapsi}\nonumber
  \begin{split}  
    S^-_\alpha\bigket{\tilde\psi_0}=\hspace{-13pt}
    \sum_{\{z_3,\dots,z_{N}\}}
    \biggl\{\sum_{z_2 (\ne \eta_\alpha)}
    &\tilde\psi_0(\eta_\alpha,z_2,z_3,\dots)
    \,S_\alpha^-\, \tilde{S}_\alpha^+\tilde{S}_{z_2}^+\\
    +\sum_{z_1 (\ne \eta_\alpha)}
    &\tilde\psi_0(z_1,\eta_\alpha,z_3,\dots)
    \,S_\alpha^-\, \tilde{S}_{z_1}^+\tilde{S}_\alpha^+\\
    &\hspace{-60pt}
    +\tilde\psi_0(\eta_\alpha,\eta_\alpha,z_3,\dots)
    \,S_\alpha^-\,(\tilde{S}_\alpha^+)^2\biggr\}
    \tilde{S}_{z_3}^+\dots\,\ket{-1}_N\\
    =\hspace{-12pt}
    \sum_{\{z_3,\dots,z_{N}\}}
    \biggl\{\sum_{z_2}
    2\hspace{20pt}&\hspace{-20pt}\tilde\psi_0(\eta_\alpha,z_2,z_3,\dots)
    \,\tilde{S}_{z_2}^+\biggr\}\tilde{S}_{z_3}^+\dots\,\ket{-1}_N\end{split}
\end{equation}
where we have used 
\begin{equation}
  \label{eq:s-s+tilde}\nonumber
  S_\alpha^-\,(\tilde{S}_\alpha^+)^n \ket{1,-1}_{\alpha}
  =n\, (\tilde{S}_\alpha^+)^{n-1}\ket{1,-1}_{\alpha}.
\end{equation}
This implies $S^-_\text{tot}\ket{\psi_0}=\sum_{\alpha=1}^N
S^-_\alpha\ket{\psi_0}=0$ if and only if $\sum_{\alpha=1}^N
\tilde\psi(\eta_\alpha,z_2,z_3,\dots)=0\ \forall\ z_2,z_3,\dots z_N$.
The Perelomov identity~\cite{perelomov71tmp156} states that this holds
for lattice sums of $e^{-\frac{\pi}{2}|\eta_\alpha|^2} G(\eta_\alpha)$
times any analytic function of $\eta_\alpha$.

\emph{Generation from filled landau levels}---Rather than proceeding
in verifying invariance properties of the non-abelian CSL state
\eqref{eq:nacsl} directly, we motivate them indirectly through
demonstrating that the state can alternatively be generated though
successive projection via the abelian CSL from the wave functions of a
filled lowest Landau level (LLL).  If we choose an auxiliary magnetic
field with a strength of one half of a Dirac flux quanta per lattice
site, the wave function for a circular droplet of $M=\frac{N}{2}$
fermions filling the LLL is given by
\begin{equation}
  \label{eq:lll}
  \phi[z_i]=\prod_{i<j}^{M}(z_i-z_j)\prod_{i=1}^{M}\,e^{-\frac{\pi}{4}|z_i|^2}.
\end{equation}
The (abelian) CSL state for spin
$S=\frac{1}{2}$~\cite{Kalmeyer-87prl2095}, which
was recently shown to be the unique and exact ground state of a local
Hamiltonian~\cite{schroeter-07prl097202}, 
\begin{equation}
  \label{eq:csl}
  \psi_0^\text{\tiny CSL}[z_i]=
  \prod_{i<j}^{M}(z_i-z_j)^2 \prod_{i=1}^{M}\,G(z_i)\,e^{-\frac{\pi}{2}|z_i|^2},
\end{equation}
where the ``particles'' $z_i$ describe spin flips $S_\alpha^+$ acting
on a ``vacuum'' state with all the spins $\dw$, and $G(\eta_\alpha)$
is as above, can be generated by Gutzwiller projection of the LLL
\eqref{eq:lll} filled once with $\up$ and once with $\dw$ spin
fermions~\cite{laughlin-90prb664,greiter02jltp1029}:
\begin{equation}
  \label{eq:cslgw}
  \ket{\psi_{0}^\text{\tiny CSL}}=\vspace{-3pt}
  \sum_{\{z,w\}}\phi[z_i]\,\phi[w_j]\,    
  c^\dagger_{z_1\up}\ldots c^\dagger_{z_M\up}\,
  c^\dagger_{w_1\dw}\ldots c^\dagger_{w_{M}\dw}\vac,
\end{equation}
where the sum extends over all partitions of the lattice sites into $z$'s
and $w$'s and the $c^\dagger$'s are fermion creation operators.  
We can rewrite the CSL state vector in terms of Schwinger bosons 
$a^\dagger$ and $b^\dagger$,
\begin{equation}
  \label{eq:cslop}\nonumber
  \ket{\psi_{0}^\text{\tiny CSL}}=
  \Psi^\text{\tiny CSL}\big[c_\up^\dagger ,c_\dw^\dagger\big]\vac=
  \Psi^\text{\tiny CSL}\big[a^\dagger ,b^\dagger\big]\vac,
\end{equation}
provided we define 
$\Psi_{0}^\text{\tiny CSL}\big[c_\up^\dagger ,c_\dw^\dagger\big]$
such that the operators are ordered according to a fixed labeling
of the lattice sites.  The non-abelian CSL state \eqref{eq:nacsl}
can thus alternatively be written as a symmetrization over two abelian
CSL states
\begin{equation}
  \label{eq:nacslop}
  \ket{\psi_0}=
  \Big(\Psi^\text{\tiny CSL}\big[a^\dagger ,b^\dagger\big]\Big)^2\vac.
\end{equation}
To verify \eqref{eq:nacslop}, use
$\frac{1}{\sqrt{2}}(a^\dagger)^n (b^\dagger)^{(2-n)}\vac =(\tilde{S}^+)^n\vac $
and 
\begin{equation}
  \label{eq:cauchy}\nonumber
  \mathcal{S}
  \hspace{-2pt}\prod_{i<j,1}^{M}\hspace{-2pt}(z_i-z_j)^2\hspace{-10pt}
  \prod_{i<j,M+1}^{2M}\hspace{-8pt}(z_i-z_j)^2
  =\text{Pf}\left(\frac{1}{z_i-z_j}\right)
  \prod_{i<j}^{2M}(z_i-z_j),
\end{equation}
where $\mathcal{S}$ indicates symmetrization.
Since the LLL states \eqref{eq:lll} are (on compact surfaces)
translationally and rotationally invariant modulo gauge
transformations in the auxiliary magnetic field, and \eqref{eq:cslgw} is
manifestly gauge covariant, both the CSL states \eqref{eq:csl} and
\eqref{eq:nacsl} are invariant under lattice transformations.  Note
that this projective construction also implies the singlet property of
the CSL states.  It can be used to formulate the CSL states on any
lattice, and to generalize them to arbitrary spin:
\begin{equation}
  \label{eq:nacslopS}
  \bigket{\psi_0^{\tiny\text{Spin}~S}}=
    \Big(\Psi^\text{\tiny CSL}\big[a^\dagger ,b^\dagger\big]\Big)^{2S}\vac.
\end{equation}  
Written in terms of (then differently) re-normalized spin flip ``particles'',
the wave function generalizes from a bosonic Pfaffian 
state for $S=1$ to bosonic Read-Rezayi states~\cite{Read-99prb8084}
for $S>1$.

\emph{Non-abelian spinon and holon excitations}---The spinon
excitations of \eqref{eq:nacsl} are analogous to the half
vortex quasiparticles of the Moore-Read quantum Hall
state~\cite{Moore-91npb362}.  For example, to create 4 $\dw$ spin
spinons at locations $\eta_1, \eta_2, \eta_3$, and $\eta_4$, we simply
insert half quantum vortices inside the Pfaffian \eqref{eq:pfaff},
which then becomes
\begin{equation}
  \label{eq:4spinons}
  \text{Pf}\left(\frac{
      (z_{i}-\eta_1)(z_{j}-\eta_2)(z_{i}-\eta_3)(z_{j}-\eta_4)
    +(i\leftrightarrow j)}
  {z_i-z_j}\right).
\end{equation}
The braiding properties of the spinons are insensitive to the spinon
spin, and are exactly those of the Moore-Read
quasiparticles~\cite{Read-00prb10267,Ivanov01prl268,Stern-04prb205338}.
The proof of the singlet property given above can be extended to show
that a pair of $\dw$ spin spinons transforms as an $S=1$ triplet
excitation, which implies that each spinon carries spin
$S=\frac{1}{2}$.  With the implicit assumption that the $S=1$ spins on
each lattice site consist of two electrons in triplet configurations,
we can create holon excitations by annihilating $\dw$ spin electrons
on sites with $\dw$ spin spinons.  The braiding properties of the
holons are equivalent to those of the spinons.

 \begin{figure}[t]
  \begin{minipage}[c]{0.45\textwidth}
    \includegraphics[width=\linewidth]{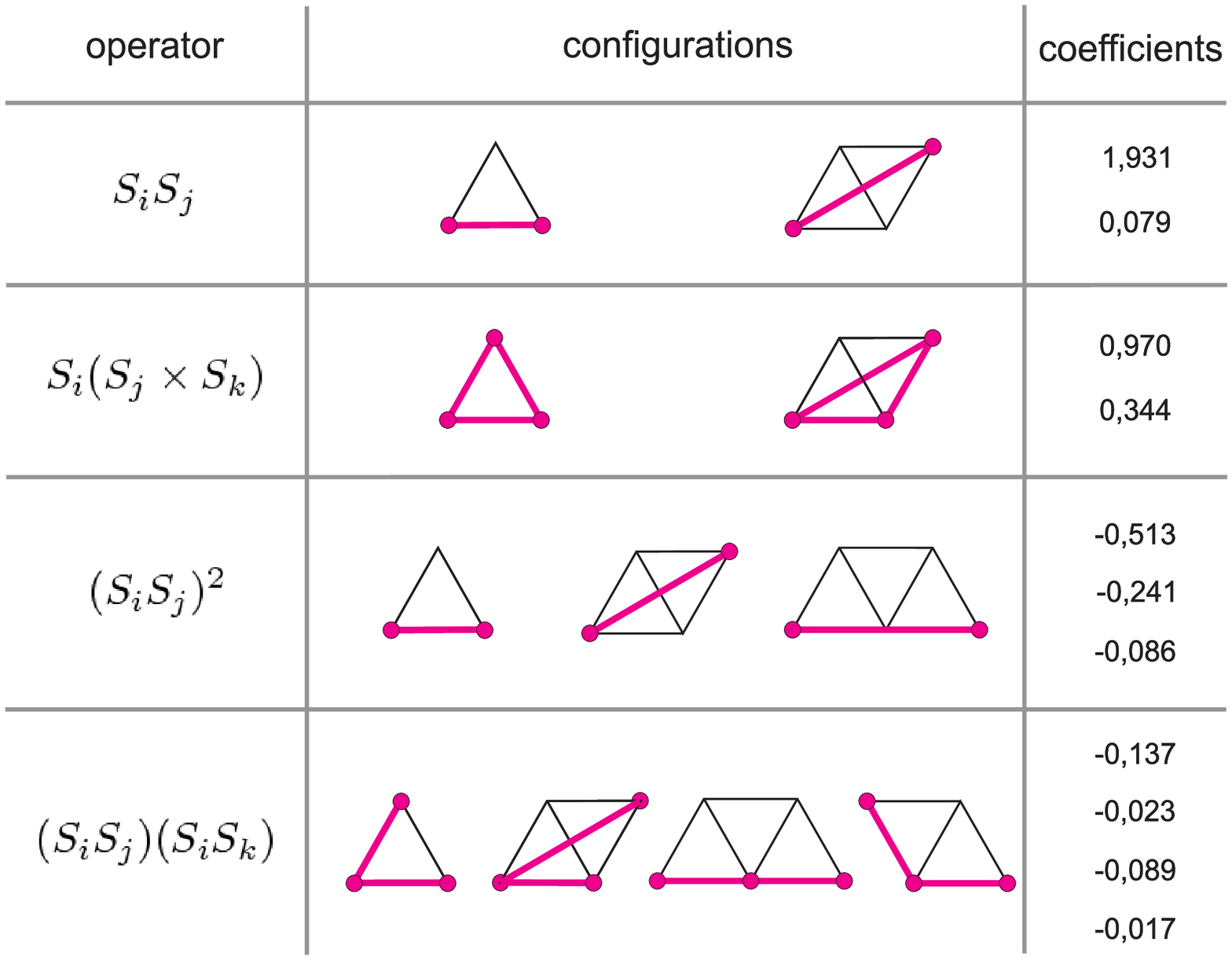}
  \end{minipage}
  \caption{(Color online) The eleven interaction terms included in our
    trial Hamiltonian with the numerically optimized coefficients (see
    text).}
  \label{fig:tab}
\vspace{-10pt}
\end{figure}

\begin{figure}
  \begin{minipage}[c]{0.45\textwidth}
    \includegraphics[width=\linewidth]{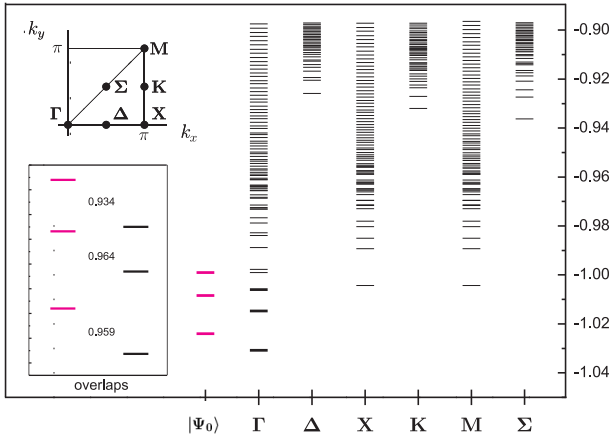}
  \end{minipage}
  \caption{(Color online) Spectral plot of our trial Hamiltonian in
    comparison with the energy expectations values for the three (in
    the infinite system topologically degenerate) Pfaffian ground
    states states at the $\Gamma$ point.  The inset shows the overlap
    of the Pfaffian states with the three lowest states of our
    Hamiltonian.}
\label{fig:spec}
\vspace{-10pt}
\end{figure}

\vspace{3pt} \emph{Model Hamiltonian}---The first question with regard
to possible applications of our state to quantum computation is
whether a state belonging to the universality class described by
\eqref{eq:nacsl} can be stabilized through a local Hamiltonian.  While
we are short of a definite answer, we have done our best to address
the question numerically.  To begin with, we have written out the
state \eqref{eq:nacsl} for an isotropic, triangular lattice with 16
sites and periodic boundary conditions, which imply a three-fold
topological degeneracy~\cite{greiter-92npb567}.  We then numerically
optimized the coefficients of a set of local spin interaction terms
(see Fig.~\ref{fig:tab}) such that the ground state of our trial
Hamiltonian is energetically closest to a suitable linear combination
of the three (in the thermodynamic limit degenerate) Pfaffian states,
which we then compare to the exact eigenstates.  As shown in
Fig.~\ref{fig:spec}, the three lowest energy eigenstates of our trial
Hamiltonian have a significant overlap with the Pfaffian states (\ie
0.959, 0.964, and 0.934 in
a fully symmetry reduced $S_{\text{tot}}^z=0$ Hilbert space with
dimension 163101), which suggests that the exact states belong to the
same universality class.
%
Note that the coefficients in Fig.~\ref{fig:tab} fall off rapidly with
the distance. Small variation of the parameters induce no sensitive
change in the overlaps, which indicates that the non-abelian CSL state
is stablilized throughout a finite region in parameter space.  Our
evidence is unfortunately not conclusive as the three CSL states are
not separated by a large gap from the remainder of the spectrum, which
indicates that the system we can access numerically is too small to
settle the question unambiguously.  
%

\emph{Experimental realization}---Recent work on polar molecules in
optical lattices~\cite{Micheli-06np341}, but in particular on
engineering 3-body interactions~\cite{Buechler-07np726}, suggests that
a realization of the non-abelian CSL proposed here might be possible
at some stage in the future.

\emph{Non-abelian spinons in general}---Efforts to understand high
$T_\text{c}$ superconductivity in terms of an RVB spin liquid have
revealed a general connection between $d$-wave superconductors and
$S=\frac{1}{2}$ spin liquids on the square
lattice~\cite{affleck-88prb745,
Lee-06rmp17}.  In
particular, a wide class of (undoped) $S=\frac{1}{2}$ spin liquids can
be obtained by Gutzwiller projection from the wave function of a
$d$-wave superconductor with suitably chosen parameters.  This
suggests a general connection between the (abelian) vortices of the
superconductor and the (abelian) spinons in the spin liquid.  If one
Gutzwiller projects a 
$d+id$ wave superconductor with suitably chosen parameter on a square
lattice, one obtains exactly the CSL state \eqref{eq:csl}.

The $p+ip$ pairing correlations in the non-abelian CSL state
\eqref{eq:nacsl} introduced above suggest a similar correspondence
between the non-abelian vortices of the superconductor and the
non-abelian spinon excitations \eqref{eq:4spinons}.  As in the abelian
case $S=\frac{1}{2}$, the P and T violation of the state appears to be
necessary for the spinon to be deconfined, but does not seem essential
to the topological properties.  
We are hence led to conjecture that
there is a general connection between $p$-wave
superfluids and $S=1$ spin liquids, in that the non-abelian braiding
properties of the vortices of the superfluid are also general
properties of the spinons in $S=1$ antiferromagnets.  
True, the spinons will only be free under special circumstances, and
the propensity to be confined will only increase with the spin $S$.
Even in an ordered antiferromagnet, however, spinons (and holons) are
the fields appropriate for describing the physics at sufficiently high
energy scales, \ie energies above the ordering temperature.

We will show elsewhere 
that the total dimension of the Hilbert space spanned by the ground
state plus all states with different numbers of spinons for the spin
liquid we propose is $3^N$, as required for a $S=1$ system with $N$
sites.  We conjecture that the non-ablelian statistics for the spinons
is not only a sufficient, but even necessary condition for the state
counting to work out consistently.  (Haldane~\cite{haldane91prl937}
has shown that the state counting for $S=\frac{1}{2}$ spin liquids
works out consistently if one assumes abelian half-fermi statistics
for the spinons.)

\emph{Conclusion}---In this work, we have constructed an \hbox{$S=1$}
CSL and argued that its spinon and holon excitations obey non-abelian
statistics.  We have used exact diagonalization studies to obtain
preliminary indication that the state can be stabilized on a $S=1$
triangular lattice.


\begin{acknowledgments}
  MG wishes to thank B.\ Paredes for a highly stimulating discussion.
  RT was supported by a PhD scholarship from the Studienstiftung des
  deutschen Volkes.
\end{acknowledgments}


\end{document}